\documentclass[aps,reprint,prd,11pt,onecolumn,notitlepage,footnote]{revtex4-2}
\usepackage[english]{babel}
\usepackage{amsmath,amssymb,graphicx,bm}
\usepackage[colorlinks=true,citecolor=blue,linkcolor=blue,urlcolor=blue]{hyperref}
\usepackage[font={small},flushleft,indent]{caption}

\begin{document}

\title{Caustics and image multiplicity of equatorial sources in Kerr spacetime}

\author{Qing-Hua Zhu}
\email{zhuqh@cqu.edu.cn}
\affiliation{Department of Physics and Chongqing Key Laboratory for Strongly
Coupled Physics, Chongqing University, Chongqing 401331, China}

\begin{abstract}
Observations of the supermassive black holes in M87 and in the Galactic Center have resolved ring-like structure at the horizon scale and have opened a regime in which the strong-field geometry can be probed directly. Motivated by the black hole images, we study sources distributed on the surface of an accretion disk confined to the equatorial plane and derive the corresponding caustic structure in Kerr spacetime. We develop a new implementation of the emitter--observer problem (EOP) in which the two-dimensional root finding for the impact parameters is reduced to two successive one-dimensional problems. With this implementation, we establish a correspondence between the image multiplicity in the EOP and the caustic structure as defined in gravitational lensing. We obtain a semi-analytic expression for the magnification and show that the caustic curves, the critical curves, and the multiplicity of the images at each order follow from a single function of one variable, the azimuthal phase of the source plane. We show how the subring is split by the caustic structure, and find that the images of an equatorial point-like source at a fixed image order lie on a common closed curve.
\end{abstract}

\maketitle

\section{Introduction}

Observations of the supermassive black holes in M87 and in the
Galactic Center have resolved ring-like structure near the horizon scale and have opened a regime
in which the strong-field geometry can be probed directly
\cite{EventHorizonTelescope:2019dse,EventHorizonTelescope:2022wkp,EventHorizonTelescope:2021srq,EventHorizonTelescope:2022xqj,EventHorizonTelescope:2022exc,EventHorizonTelescope:2022urf,Genzel:2024vou}. The next-generation very long baseline interferometers
aim at an order-of-magnitude improvement in baseline length, which would make it possible to resolve structures
at the scale of the photon ring \cite{Johnson:2023ynn,Johnson:2024ttr,Kim:2023hnr,Valtonen:2025ovs}.
Among the targets of these instruments, the secondary and higher-order images
play a special role. A higher-order image is formed by light that has
wound around the black hole one or more times before escaping to the observer. The
existence of such an image is independent of the emission mechanism and of the source
configuration, and is therefore regarded as a clean probe of the
spacetime geometry \cite{Narayan:2023ivq,Vagnozzi:2022moj,EventHorizonTelescope:2022xqj,Wielgus:2021peu,Hadar:2020fda,Narayan:2019imo,Liu:2025wwq,Wei:2026zwu,Ren:2026dnv,Wan:2026lca}. 

For a ring-like source such as the accretion disk around a black hole, the image
order is unambiguously counted by the number of half-orbits $n$ that the polar angle $\theta$ completes
\cite{Luminet:1979nyg,Gralla:2019xty,Gralla:2019ceu}. The case $n=0$ gives the
direct image, and the images with $n\geq1$ form the so-called subring or subimage
\cite{Johnson:2019ljv,Wong:2024gph}. As $n$ grows, the subring becomes
narrower and approaches the sharp photon ring. This picture does not depend on
whether the black hole rotates or not. By contrast, for point-like sources such as a distant star,
the situation is different. Strong gravitational lensing shows that the caustic
structure of a rotating black hole splits the images further, beyond the classification by
the image order $n$ alone
\cite{1994ApJ...421...46R,Bozza:2002af,Bozza:2008mi,Sereno:2007gd,Schneider:1992bmb,Saha:2024axf,Bozza:2010xqn,Feleppa:2026emm,Bozza:2002zj,Perlick:2004tq,Perlick:2003vg,Virbhadra:1999nm,Cunha:2018acu}.
A gap therefore separates the ring-like source from the point-like source. For a
ring-like source, it is not
clear whether the caustics of a rotating black hole affect its image. This question is one
of the motivations for this study \footnote{A similar question on image
order degeneracy has been raised and discussed in Refs.~\cite{Zhou:2024dbc,Wong:2024gph}.}. 

A point-like source confined to the equatorial plane is of interest for the direct
observation of a supermassive black hole. It is idealized as the ``hot spot'' model,
in which the flare emission comes from a compact, optically thin source that follows
a geodesic of the background spacetime and emits synchrotron radiation
\cite{1992A257594B,1994ApJ...425...63B,Broderick:2005jj,Broderick:2005my,Li:2014fza,Matsumoto:2020wul,Zhu:2024vxw,Xie:2025skg}.
This model has now been confronted with the astrometry, the polarimetry, and the light
curves of the Sgr A* flares
\cite{Yfantis:2023wsp,Yfantis:2024eab,Levis:2023tpb,Antonopoulou:2024qco,GRAVITY:2018sef,GRAVITY:2020lpa,GRAVITY:2023avo,GRAVITY:2020xcu,GRAVITY:2020hwn,Xie:2025skg,Wang:2025btn,Qin:2025zdv}. 

For a source that occupies a small portion of the accretion flow, the computationally
efficient way to trace light rays for imaging is to solve an
emitter--observer problem (EOP) \cite{1993A&A...272..355V,Beckwith:2004ae}. Given an
emitter position $\mathbf{x}_\text{s}$ and an observer position
$\mathbf{x}_\text{o}$, one finds all light rays that connect the two. The spirit and the term of the emitter--observer problem go back to Viergutz \cite{1993A&A...272..355V}.
Public implementations of EOP codes are also available
\cite{Yang:2013yoa,Yang:2013wya}. The approach was revisited recently and named
``forward ray tracing'' \cite{Zhou:2024dbc}. For a rotating black hole,
the implementation of the EOP turns the solution of the geodesic equations into a
root-finding procedure in a two-dimensional space
\cite{1993A&A...272..355V,Beckwith:2004ae,Zhou:2024dbc}, or even in a
three-dimensional space \cite{Yang:2013yoa}. The roots are multiple, and
they correspond to the different light rays that connect the same emitter and observer.
The complexity of a given implementation therefore lies in the indices used to label the different types of light rays.

Strong gravitational lensing provides the standard formalism for describing the multiplicity of the
images, namely the caustic structure
\cite{Schneider:1992bmb,Saha:2024axf,Bozza:2010xqn,Bozza:2008mi}. The caustic curves
are determined by the vanishing of the Jacobian determinant of the lens equation, and
they are defined with respect to a source plane, usually the distant celestial sphere
in terms of the angular coordinates $(\theta_\text{s},\phi_\text{s})$. In an EOP
solver, by contrast, the multiplicity of the images is not an additional ingredient.
It is already contained in the existence of multiple roots for the impact parameters. In other words, the
existing EOP studies appear not to rely on the language of caustics \cite{1993A&A...272..355V,Yang:2013yoa,Zhou:2024dbc}. However, the EOP and the caustic structure both describe the multiplicity of the images, and there must be an intrinsic relation between them. 

In this work, we develop a new implementation of the EOP that
is suited to sources on the equatorial plane, in which the two-dimensional
root finding of Ref.~\cite{1993A&A...272..355V,Zhou:2024dbc} is replaced by two
successive one-dimensional problems. We consider a lensing system whose source plane is the
equatorial plane \cite{1992A257594B,1994ApJ...425...63B,Zhu:2026jom}, and we allow the source to lie arbitrarily close to the
horizon of a Kerr black hole. We aim to establish the correspondence between the
image multiplicity in the EOP and the caustic structure of strong
lensing, and we apply the resulting formalism to three situations. The first is
the construction of the caustic and critical curves on the source plane
$(r_\text{s},\phi_\text{s})$. The second is a complete count of the
images at each order. The third is a study of the
magnification of these images in the presence of the full caustic structure.

The paper is organized as follows. In Sec.~\ref{sec:kerr}, we recall the Kerr
metric, the integral form of the null geodesic equations, and the branches and
domains of the impact parameters. In Sec.~\ref{sec:eop}, we present the new
implementation of the EOP, formulate the lens mapping for an equatorial
source, and derive the semi-analytic magnification. In
Sec.~\ref{sec:caustics}, we study the morphology of the azimuthal phase, the
caustic and critical curves, the magnification, and the multiplicity of the
images. Sec.~\ref{sec:conclusions} collects our conclusions and discusses
the limitations of the approach.

\section{Kerr metric and geodesic equations}
\label{sec:kerr}

In this section, we summarize the transfer equations of Kerr spacetime and the
classification of the null geodesics by their radial turning point and by the number
of half-orbits. We also determine the domain of the impact parameters for each class
of geodesics. This analysis is a prerequisite for the implementation of the EOP, as
 indicated by the pioneering studies of Refs.~\cite{1993A&A...272..355V,Beckwith:2004ae}.

\subsection{Null geodesics and the image plane}

We work in Boyer--Lindquist coordinates $(t,r,\theta,\phi)$ and in units with
$G=c=1$. The Kerr metric is
\begin{equation}
\mathrm{d}s^2=-\frac{\Delta}{\Sigma}\left(\mathrm{d}t-a\sin^2\theta\,\mathrm{d}\phi\right)^2
+\frac{\sin^2\theta}{\Sigma}\left[a\,\mathrm{d}t+(r^2+a^2)\mathrm{d}\phi\right]^2
+\frac{\Sigma}{\Delta}\mathrm{d}r^2+\Sigma\,\mathrm{d}\theta^2,
\label{eq:metric}
\end{equation}
where
\begin{equation}
\Delta=r^2-2Mr+a^2,\qquad \Sigma=r^2+a^2\cos^2\theta.
\label{eq:DeltaSigma}
\end{equation}
$M$ is the mass, and $a=J/M$ the spin parameter. The transfer equations that connect an emission point $x_\text{s}=(t_\text{s},r_\text{s},\theta_\text{s},\phi_\text{s})$ with an
observer $x_\text{o}=(t_\text{o},r_\text{o},\theta_\text{o},\phi_\text{o})$
can be written as \cite{Misner:1973prb,Chandrasekhar:1985kt}
\begin{subequations}
  \begin{align}
0&=G_{\theta}(\theta_\text{o},\theta_\text{s})-I_r(r_\text{o},r_\text{s}),
\label{eq:geod1}\\
\phi_\text{o}-\phi_\text{s}&=a I_{\phi}(r_\text{o},r_\text{s})
+\lambda G_{\phi}(\theta_\text{o},\theta_\text{s}),
\label{eq:geod2}\\
t_\text{o} - t_\text{s} & =  I_t (r_\text{o}, r_\text{s}) + a^2   G_t (\theta_\text{o}, \theta_\text{s}),
\label{eq:geod3}
\end{align}
\end{subequations}
with the transfer functions
\begin{subequations}
\label{eq:transfers}
\begin{align}
I_r&=\int_{r_\text{s}}^{r_\text{o}}\frac{\mathrm{d}r}{\sqrt{R(r)}}~,
&&G_{\theta}=\int_{\theta_\text{s}}^{\theta_\text{o}}\frac{\mathrm{d}\theta}{\sqrt{\Theta(\theta)}}~, \label{eq:Ir}\\
I_{\phi}&=\int_{r_\text{s}}^{r_\text{o}} \mathrm{d}r \left\{\frac{2Mr-a\lambda}{\Delta\sqrt{R(r)}} \right\}~,
&&G_{\phi}=\int_{\theta_\text{s}}^{\theta_\text{o}}\mathrm{d}\theta \left\{\frac{\csc^2\theta}{\sqrt{\Theta(\theta)}}\right\}~, \label{eq:Iphi}\\
I_t & =  \int_{r_\text{s}}^{r_\text{o}} \mathrm{d} r \left\{ \frac{r^2 \Delta + 2 M
  r (r^2 + a^2 - a \lambda)}{\Delta \sqrt{R (r)}} \right\},
&&G_t  =  \int^{\theta_\text{o}}_{\theta_\text{s}} \mathrm{d} \theta \left\{ \frac{\cos^2
  \theta}{\sqrt{\Theta (\theta)}} \right\}~. \label{eq:It}
\end{align}
\end{subequations}
In this study, we use the two radial and two angular transfer functions of
Eq.~\eqref{eq:transfers}.  The functions $I_t$ and $G_t$ are not needed for the images. The radial and angular potentials in Eq.~\eqref{eq:transfers} are
\begin{equation}
R(r)=\left(r^2+a^2-a\lambda\right)^2-\Delta\kappa,
\qquad
\Theta(\theta)=\kappa-\left(\lambda-a\sin^2\theta\right)^2\csc^2\theta.
\label{eq:potentials}
\end{equation}
The constants $\kappa$ and $\lambda$ are the impact parameters that label the geodesic \cite{Grenzebach:2014fha,Zhu:2023kei}. For an observer at
large distance, the image plane is parametrized by polar coordinates
$(\rho,\varphi)$ through \cite{Gralla:2019drh}
\begin{equation}
\lambda=-\rho\sin\theta_\text{o}\cos\varphi,\qquad
\kappa=\left(\rho\cos\varphi+a\sin\theta_\text{o}\right)^2+\rho^2\sin^2\varphi,
\label{eq:screen}
\end{equation}
so that a ray is specified by the position $(\rho,\varphi)$ at which it reaches the observer's screen, also known as the image plane.

\subsection{Branches and domains of the impact parameters}
\label{sec:branch}

Both potentials of Eq.~\eqref{eq:potentials} may change sign along a geodesic,
and each sign change introduces a branch in the integrals of
Eq.~\eqref{eq:transfers}. It is convenient to write them in the factorized form
\begin{equation}
R(r)=\prod_{i=1}^{4}(r-r_i),\qquad
\Theta(\theta)=\frac{a^2}{\sin^2\theta}\left(u_+-\sin^2\theta\right)\left(\sin^2\theta-u_-\right),
\label{eq:factored}
\end{equation}
where $r_{1,2,3,4}$ are the roots of $R(r)=0$, given in
closed form in Appendix~\ref{app:integrals}, and $u_\pm$ are the two roots of
$\Theta=0$.

For the radial motion, the four roots can all be real, or two can be real with
the remaining pair complex conjugate, or the four roots can form two complex
conjugate pairs. In the last case, $R(r)$ has no real root and never changes
sign, so the light ray has no radial turning point. This case plays no role in the
branch structure discussed below. When all four roots are real and $r_\text{s}\geq r_4$, the largest root
$r_4$ is a radial turning point, and the radial integral is evaluated as
\begin{equation}
I_\ast=\begin{cases}
\displaystyle\int_{r_\text{s}}^{r_\text{o}}\mathrm{d}r\,\{\ldots\}, & \text{without turning point},\\[2mm]
\displaystyle\left[\int_{r_4}^{r_\text{s}}+\int_{r_4}^{r_\text{o}}\right]\mathrm{d}r\,\{\ldots\}, & \text{with turning point},
\end{cases}
\label{eq:turning}
\end{equation}
where $\ast$ stands for $r$ or $\phi$. We denote by $I_\ast^\text{in}$
and $I_\ast^\text{out}$ the branches without and with a radial turning point,
respectively. The two branches coincide when $r_\text{s}=r_4$, which is
therefore the critical condition that separates the light rays that turn at $r_4$ from
those that do not. Imposing $R(r_\text{s})=0$ and using
Eq.~\eqref{eq:screen}, this condition becomes
\begin{equation}
T_r(\rho,\varphi)\equiv\left(1-\frac{a^2\sin^2\theta_\text{o}\cos^2\varphi}{\Delta_\text{s}}\right)\rho^2
-\frac{4Mr_\text{s}a\sin\theta_\text{o}\cos\varphi}{\Delta_\text{s}}\rho
+a^2\sin^2\theta_\text{o}-\frac{(r_\text{s}^2+a^2)^2}{\Delta_\text{s}}=0,
\label{eq:Tr}
\end{equation}
with $\Delta_\ast\equiv\Delta(r_\ast)$. A light ray has a radial turning point when
$T_r(\rho,\varphi)>0$ and does not have one otherwise. In addition, a real root
$r_4$ outside the horizon exists only for screen positions outside the shadow, whose
boundary is determined by the unstable photon orbits, $R(r)=R'(r)=0$ \cite{Bardeen:1973tla}. Writing $r_\text{c}$ for
the radius of a photon sphere and eliminating $\kappa$ and $\lambda$ with Eq.~\eqref{eq:screen},
we obtain the shadow boundary as a parametric curve, namely,
\begin{subequations}
  \begin{align}
\rho^2&=\frac{16r_\text{c}^2\Delta_\text{c}}{(\Delta'_\text{c})^2}
-\frac{8r_\text{c}\Delta_\text{c}}{\Delta'_\text{c}}
+2\left(r_\text{c}^2+a^2\right)-a^2\sin^2\theta_\text{o},
\label{eq:shadow1}\\
\rho\cos\varphi&=\frac{1}{\sin\theta_\text{o}}\left(\frac{4r_\text{c}\Delta_\text{c}}{\Delta'_\text{c}}-\frac{r_\text{c}^2}{a}-a\right),
\label{eq:shadow2}
\end{align}
\end{subequations}
where $\Delta'_\text{c}\equiv\mathrm{d}\Delta/\mathrm{d}r|_{r_\text{c}}$ and
$r_\text{c}$ runs over the photon sphere. Horizon-scale observations have used this
boundary to estimate the parameters of the black hole and to constrain alternatives to
the Kerr metric
\cite{Kumar:2018ple,Tian:2019yhn,EventHorizonTelescope:2020qrl,EventHorizonTelescope:2021dqv,Vagnozzi:2022moj,Liu:2025wwq}.

For the angular motion, we consider rays that can reach the equatorial plane.
This requires $u_+>1$ and $0<u_-<1$. Rays that cannot reach the equatorial
plane are the vortical geodesics classified in Refs.~\cite{Bardeen:1973tla,Gralla:2019drh,Gralla:2019ceu,Maliborski:2025ovf}. Since
$u_-\in(0,1)$, the polar angle oscillates between
$\theta_-=\arcsin\sqrt{u_-}$ and $\theta_+=\pi-\arcsin\sqrt{u_-}$, so that the
angular integrals depend on the number of half-orbits. With the
convention $\mathrm{sign}(\varphi)=\mathrm{sign}\,[(\mathrm{d}\theta/\mathrm{d}\tau)|_{\mathbf{x}=\mathbf{x}_\text{o}}]$,
the angular integrals read
\begin{equation}
G_\ast=\mathrm{sign}(\varphi)\left[\int_{\theta_-}^{\theta_\text{o}}
-(-1)^n\int_{\theta_-}^{\theta_\text{s}}\right]\mathrm{d}\theta\,\{\ldots\}
+\left(n-\frac{1-(-1)^n}{2}\mathrm{sign}(\varphi)\right)
\int_{\theta_-}^{\theta_+}\mathrm{d}\theta\,\{\ldots\},
\label{eq:Gn}
\end{equation}
where  $\ast$ stands for $\theta$ or $\phi$, and $n$ is the number of half-orbits. We denote $G_\ast^{(n)}$ for the branch
$n$. For an equatorial source, which is the case of interest here, the light rays are
normal geodesics rather than vortical ones. The boundary
between these two classes is
\begin{equation}
T_{\theta}(\rho,\varphi)\equiv\rho^2\left(1-\sin^2\theta_\text{o}\cos^2\varphi\right)-a^2\cos^2\theta_\text{o}=0,
\label{eq:Ttheta}
\end{equation}
and in this study we restrict ourselves to $T_\theta(\rho,\varphi)>0$.

\begin{figure}[htbp]
\includegraphics[width=\linewidth]{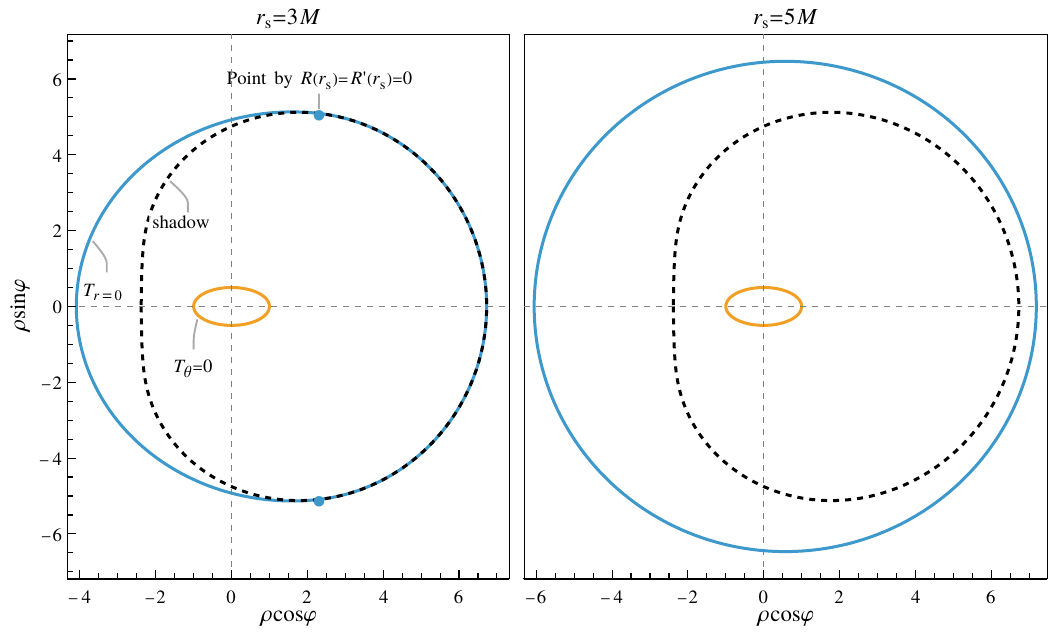}
\caption{\label{fig:boundary}Characteristic curves in the image plane: the
solid blue curve is the radial boundary $T_r(\rho,\varphi)=0$ of
Eq.~\eqref{eq:Tr} for two source radii, $r_\text{s}=3M$ (left) and
$r_\text{s}=5M$ (right). Inside this curve the light rays have no radial
turning point. The dashed
curve is the shadow boundary, obtained from $R(r)=R'(r)=0$ and given by
Eqs.~\eqref{eq:shadow1} and~\eqref{eq:shadow2}. The small orange oval is
the polar boundary $T_\theta(\rho,\varphi)=0$ of Eq.~\eqref{eq:Ttheta}. Only
the region $T_\theta>0$, in which light rays can reach the equatorial plane, is
considered in this work. In these schematic diagrams, we set $a=0.999M$ and $\theta_\text{o}=\pi/3$.}
\end{figure}

In Fig.~\ref{fig:boundary}, we
illustrate the characteristic curves of Eqs.~\eqref{eq:Tr},
\eqref{eq:shadow1}, \eqref{eq:shadow2}, and \eqref{eq:Ttheta} in the image
plane. The three curves delimit the domain on which the
transfer functions are defined, and they fix the region in which the branch of the
radial integral, $I_r^\text{in}$ or $I_r^\text{out}$, has to be chosen. The polar
boundary of Eq.~\eqref{eq:Ttheta} is a small oval around the centre of the
image plane. Inside that oval, the light rays cannot reach the equatorial plane, so the region
relevant for an equatorial source is the part of the image plane outside the oval.

\section{The emitter--observer problem, lens equation, and caustic structure}
\label{sec:eop}

The EOP is the procedure of finding all light rays that connect a given
emitter $\mathbf{x}_\text{s}$ with a given observer $\mathbf{x}_\text{o}$ in the
presence of gravitational deflection. For the Kerr black hole, it was first formulated by Viergutz
\cite{1993A&A...272..355V} by means of a two-dimensional root finding in the image
plane. In this section, we present an implementation in which the
root finding reduces to a sequence of one-dimensional problems. We use it
to formulate the lens equation and the caustic structure for a source on the
equatorial plane.

\subsection{A one-dimensional implementation of the emitter--observer problem}
\label{sec:eopnew}

The image order needed to solve the EOP is not simply the integer $n$ that
appears in $G_\ast^{(n)}$ in Eq.~\eqref{eq:Gn}. For sources with
$\sin\theta_\text{s}>\sin\theta_\text{o}$, which is the regime considered
here, it has to be defined through the number $n$ of incremental half-orbits
taken by $\theta_\text{s}$ towards $\theta_\text{o}$, as adopted in
Ref.~\cite{Zhu:2024vxw}. Explicitly, we have
\begin{align}
G_\ast^{(\text{eop},n)}(\theta_\text{o},\theta_\text{s};\rho,\varphi) &= \mathcal{H} (\text{sign}   (\theta_\text{o} - \theta_\text{s}) \varphi)
  G_{\ast}^{(n)} + \mathcal{H} (\text{sign}   (\theta_\text{s} - \theta_\text{o}) \varphi)
   G_{\ast}^{ (n + 1)} \nonumber \\
&=
\left(n+\mathcal{H}\!\left[\mathrm{sign}(\theta_\text{s}-\theta_\text{o})\varphi\right]
-\frac{\mathrm{sign}(\varphi)+\mathrm{sign}(\theta_\text{s}-\theta_\text{o})(-1)^n}{2}\right)
\int_{\theta_-}^{\theta_+}\mathrm{d}\theta\,\{\ldots\}\nonumber\\
&\quad+\mathrm{sign}(\varphi)\int_{\theta_-}^{\theta_\text{o}}\mathrm{d}\theta\,\{\ldots\}
+\mathrm{sign}(\theta_\text{s}-\theta_\text{o})(-1)^n\int_{\theta_-}^{\theta_\text{s}}\mathrm{d}\theta\,\{\ldots\},
\label{eq:Gop}
\end{align}
where $\mathcal{H}(x)$ is the Heaviside step function. In Eq.~\eqref{eq:Gop}, the branch index is counted from the source side \cite{Zhu:2024vxw}. This is also the counting that reproduces the image order in the caustic structure below.

Our implementation proceeds in two steps. First, the screen radius $\rho$ is
determined from Eq.~\eqref{eq:geod1} at fixed screen azimuth $\varphi$, namely,
\begin{equation}
I_r(r_\text{o},r_\text{s};\rho,\varphi)=G_{\theta}^{(\text{eop},n)}(\theta_\text{o},\theta_\text{s};\rho,\varphi),
\label{eq:rhoeq}
\end{equation}
whose solution we write as
\begin{equation}
\rho=\mathrm{P}^{(n)}(r_\text{s},\theta_\text{s},r_\text{o},\theta_\text{o};\varphi).
\label{eq:P}
\end{equation}
For $\sin\theta_\text{s}>\sin\theta_\text{o}$ and given
$\mathbf{x}_\text{o}$, $\mathbf{x}_\text{s}$, and $n$,
Eq.~\eqref{eq:rhoeq} has a single solution. Here, we verify the monotonicity of $P^{(n)}$ numerically and do not attempt a rigorous proof. The monotonicity is what allows the operator $\mathrm{P}^{(n)}$ of Eq.~\eqref{eq:P} to be treated as a function, which we use repeatedly in Secs.~\ref{sec:lens} and~\ref{sec:caustics}.

\begin{figure}[htbp]
\includegraphics[width=\linewidth]{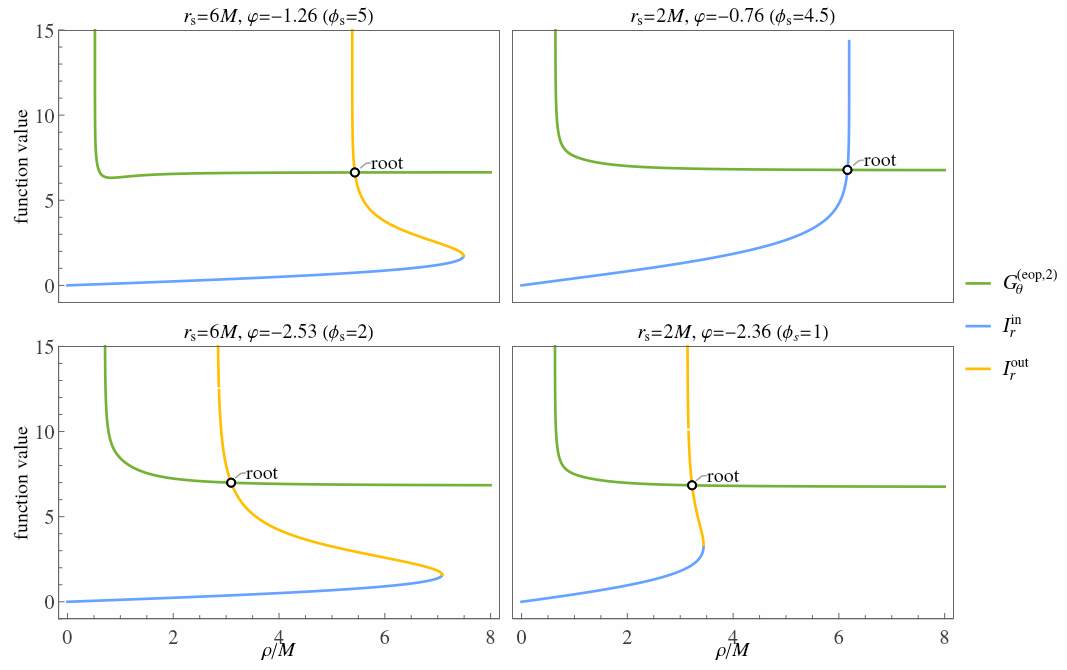}
\caption{\label{fig:rho} The schematic diagrams show the procedure of root finding for Eq.~\eqref{eq:rhoeq} at selected values of $r_\text{s}$ and $\varphi$. In these cases, we set image order $n=2$, $r_\text{o}=20M$, and black hole spin $a=0.999M$. Each intersection, marked by a dot, is a
root of Eq.~\eqref{eq:rhoeq}.}
\end{figure}

In Fig.~\ref{fig:rho}, we display several examples of the solution of Eq.~\eqref{eq:rhoeq}. The branch of $I_r$ is selected by whether the light ray possesses a radial turning point. The same root-finding procedure can be used for a spherical black hole \cite{Zhu:2024vxw,Igata:2025glk}. The difference is that here the transfer functions have to be evaluated separately at each screen azimuth $\varphi$. Consequently, the curves in the upper and lower panels of Fig.~\ref{fig:rho} differ even when the source radius is unchanged.

Second, the screen azimuth $\varphi$ follows from Eq.~\eqref{eq:geod2} after substituting the operator $\mathrm{P}^{(n)}$ of Eq.~\eqref{eq:P}, namely,
\begin{equation}
\phi_\text{o}-\phi_\text{s}
=\Big[aI_{\phi}(r_\text{o},r_\text{s};\rho,\varphi)
-\rho\sin\theta_\text{o}\cos\varphi\,G_{\phi}^{(\text{eop},n)}(\theta_\text{o},\theta_\text{s};\rho,\varphi)\Big]_{\rho=\mathrm{P}^{(n)}(r_\text{s},\theta_\text{s},r_\text{o},\theta_\text{o};\varphi)},
\label{eq:phieq}
\end{equation}
which is an equation for $\varphi$ alone, with the emitter and observer
positions and the image order fixed. Solving it numerically yields the
azimuths. With Eqs.~\eqref{eq:P} and \eqref{eq:phieq}, we obtain
the solution of the EOP, namely,
\begin{equation}
\begin{cases}
\rho=\mathrm{P}^{(n)}\!\left(r_\text{s},\theta_\text{s},r_\text{o},\theta_\text{o};\Psi^{(n)}(r_\text{s},\theta_\text{s},\phi_\text{s},r_\text{o},\theta_\text{o},\phi_\text{o})\right),\\[1mm]
\varphi=\Psi^{(n)}(r_\text{s},\theta_\text{s},\phi_\text{s},r_\text{o},\theta_\text{o},\phi_\text{o}),
\end{cases}
\label{eq:EOP}
\end{equation}
where $\Psi^{(n)}$ denotes the solution of Eq.~\eqref{eq:phieq} for $\varphi$.

\begin{figure}[htbp]
\includegraphics[width=\linewidth]{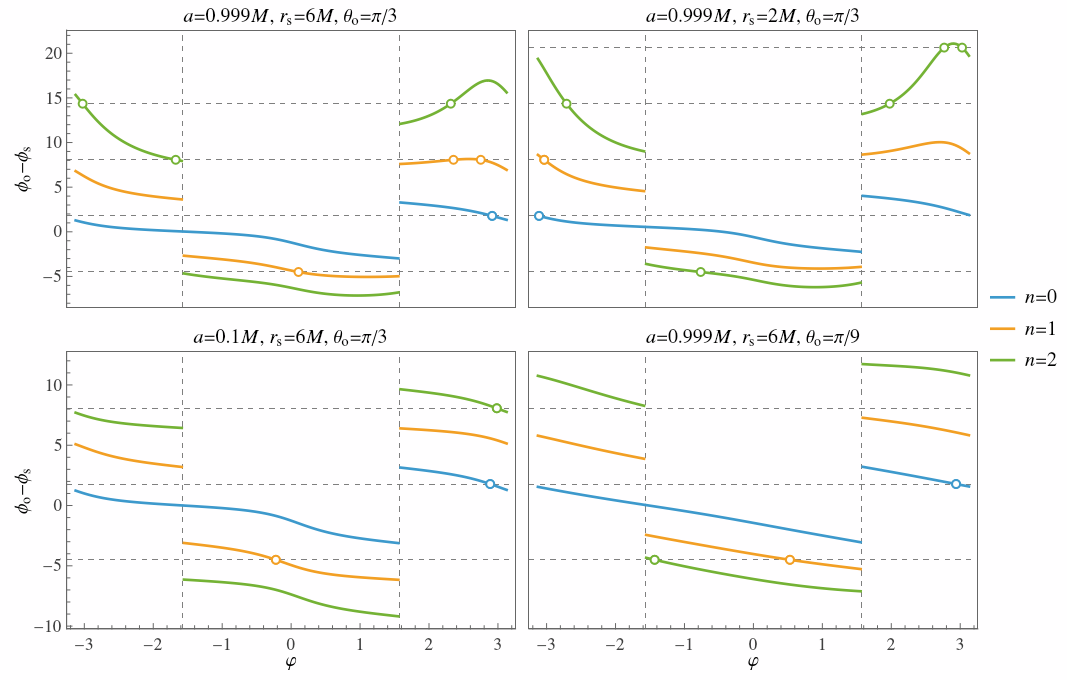}
\caption{\label{fig:phi}Right-hand side of Eq.~\eqref{eq:phieq} as a function
of the screen azimuth $\varphi$, for image orders $n=0$, $1$, and $2$. The horizontal dashed lines mark
$\phi_\text{o}-\phi_\text{s}=2\pi k$ with $k\in\mathbb{Z}$. Every crossing of
a branch with one of these levels is a solution of Eq.~\eqref{eq:phieq}, that
is, an image. In these cases, we set $r_\text{o}=20M$ and $\phi_\text{s}=4.5$.}
\end{figure}

Contrary to the case of a Schwarzschild black hole \cite{Zhu:2024vxw},
Eq.~\eqref{eq:phieq} may possess several solutions for $\varphi$ at a given
image order. 
 This multiplicity is shown in Fig.~\ref{fig:phi}.
Each branch is smooth away from $\varphi=\pm\pi/2$, where it jumps by $2\pi$.
The jump is a property of the elliptic integrals entering $G_\theta$ and not of the
geometry. It can be removed by the continuation constructed in
Appendix~\ref{app:continuation}. Every crossing of a branch with one of the
horizontal levels is a solution of Eq.~\eqref{eq:phieq} and therefore an image. For $n=0$,
the branch crosses the levels once in the whole range. For $n=1$ and $n=2$,
the branch winds more than once and the number of crossings grows accordingly. The
multiplicity is a property of the winding of a one-dimensional function. It
can therefore be counted without a two-dimensional search \cite{1993A&A...272..355V,Zhou:2024dbc}.

\begin{figure}[htbp]
\includegraphics[width=1\linewidth]{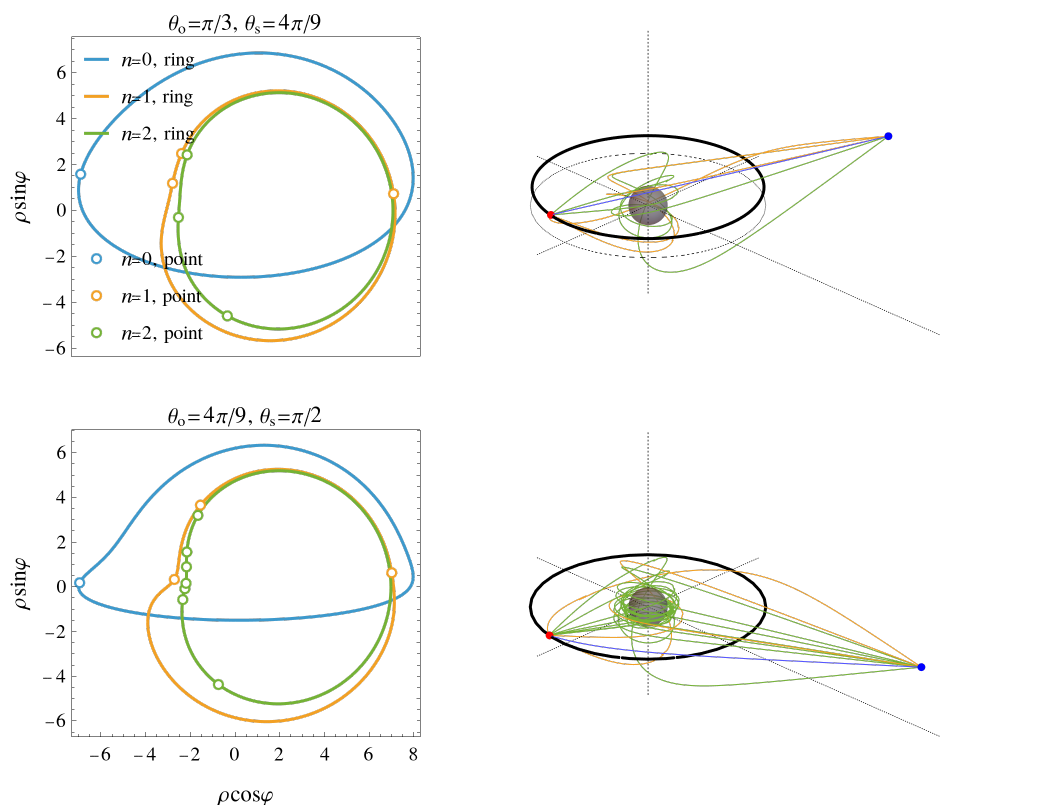}
\caption{\label{fig:image}
 Left panels: the images of a ring-like source (coloured closed curves) and of a point source (coloured open dots) in the image plane, at image orders $n=0$, $1$, and $2$. Right panels: the corresponding light paths from the point source (red) to the observer (blue) in three-dimensional space. The ring-like sources around the black hole are drawn in black; they are mapped onto the coloured closed curves in the image plane. In these cases, we set $r_\text{s}=6M$, $r_\text{o}=20M$ and $\phi_\text{s}=4.5$.}
\end{figure}

The corresponding images of the point source of Fig.~\ref{fig:phi}, and the light
paths that produce them, are displayed in Fig.~\ref{fig:image}. The right panels show these light paths, which wind around the black hole a number of times set by the image order. For a given image order, the images of the point source lie on one and the same closed curve of the image plane, as shown in the left panels. This property is
specific to the regime $\sin\theta_\text{s}>\sin\theta_\text{o}$, in which Eq.~\eqref{eq:rhoeq} has a single solution. Outside this regime, the multiple images of the same order need not share a common curve. The complexity may increase, as suggested in Ref.~\cite{Feleppa:2026emm}.

In the following, we show that these multiple solutions can be linked to the caustics
of strong gravitational lensing. The formalism of Eqs.~\eqref{eq:rhoeq} and \eqref{eq:phieq} is well suited to a comprehensive analysis of the caustic structure.

\subsection{Formulating the caustic structure of an equatorial source}
\label{sec:lens}

The conventional lens equation establishes a map between an angular source
plane $(\theta_\text{s},\phi_\text{s})$ at large distance and the image plane
$(\rho,\varphi)$. Here, we address instead the surface of an accretion disk
close to the black hole, and we focus on the caustic structure of a source on
the equatorial plane, $\theta_\text{s}=\pi/2$. The lens equation then maps the
image plane $(\rho,\varphi)$ onto the source plane
 $(r_\text{s},\phi_\text{s})$, and it reads
\begin{equation}
\begin{cases}
r_\text{s}=R^{(n)}_{r_\text{o},\theta_\text{o},\theta_\text{s}}(\rho,\varphi),\\[1mm]
\phi_\text{s}=\Phi^{(n)}_{r_\text{o},\theta_\text{o},\theta_\text{s},\phi_\text{o}}(\rho,\varphi),
\end{cases}
\label{eq:lens}
\end{equation}
where
\begin{subequations}
    \begin{align}
R^{(n)}_{r_\text{o},\theta_\text{o},\theta_\text{s}}(\rho,\varphi)
&\equiv I_r^{-1}\!\left(G_\theta^{(\text{eop},n)}(\theta_\text{o},\theta_\text{s};\rho,\varphi);r_\text{o}\right),
\label{eq:Rn}\\
\Phi^{(n)}_{r_\text{o},\theta_\text{o},\theta_\text{s},\phi_\text{o}}(\rho,\varphi)
&\equiv\phi_\text{o}-\Big[aI_{\phi}(r_\text{o},r_\text{s};\rho,\varphi)
-\rho\sin\theta_\text{o}\cos\varphi\,G_\phi^{(\text{eop},n)}(\theta_\text{o},\theta_\text{s};\rho,\varphi)\Big]_{r_\text{s}=R^{(n)}_{r_\text{o},\theta_\text{o},\theta_\text{s}}(\rho,\varphi)}.
\label{eq:Phin}
\end{align} \label{eq:RPhi}
\end{subequations}
The Jacobian of this map defines the magnification. It is given by $\Sigma$, a
quantity proportional to the dimensionless magnification factor $\mu$, namely,
\begin{equation}
\frac{1}{\Sigma}=\frac{r_\text{s}}{\rho}\left|\frac{\partial(r_\text{s},\phi_\text{s})}{\partial(\rho,\varphi)}\right|
\equiv\frac{r_\text{s}}{\rho}\left|\frac{\partial\left(R^{(n)}_{r_\text{o},\theta_\text{o},\theta_\text{s}},\Phi^{(n)}_{r_\text{o},\theta_\text{o},\theta_\text{s},\phi_\text{o}}\right)}{\partial(\rho,\varphi)}\right|,
\label{eq:invSigma}
\end{equation}
and the caustics are the locations at which $\Sigma\to\infty$ on the source plane.

To evaluate $\Sigma$, we substitute the operator $\mathrm{P}^{(n)}$ into the
lens equation of Eq.~\eqref{eq:lens}. The first equation of Eq.~\eqref{eq:lens} then
becomes an identity,
\begin{equation}
r_\text{s}=R^{(n)}_{r_\text{o},\theta_\text{o},\theta_\text{s}}\!\left(\mathrm{P}^{(n)}(r_\text{s},\theta_\text{s},r_\text{o},\theta_\text{o};\varphi),\varphi\right)=r_\text{s},
\label{eq:identity}
\end{equation}
while the second equation becomes
\begin{equation}
\phi_\text{s}=\Phi^{(n)}_{r_\text{o},\theta_\text{o},\theta_\text{s},\phi_\text{o}}\!\left(\mathrm{P}^{(n)}(r_\text{s},\theta_\text{s},r_\text{o},\theta_\text{o};\varphi),\varphi\right).
\label{eq:phimaster}
\end{equation}
Eq.~\eqref{eq:phimaster} is the same equation that is solved in the EOP [Eq.~\eqref{eq:phieq}]. Differentiating Eqs.~\eqref{eq:identity} and
\eqref{eq:phimaster} gives
\begin{subequations}
    \begin{align}
\mathrm{d}r_\text{s}&=\frac{\partial R^{(n)}}{\partial\rho}\frac{\partial\mathrm{P}^{(n)}}{\partial r_\text{s}}\,\mathrm{d}r_\text{s}
+\left(\frac{\partial R^{(n)}}{\partial\rho}\frac{\partial\mathrm{P}^{(n)}}{\partial\varphi}
+\frac{\partial R^{(n)}}{\partial\varphi}\right)\mathrm{d}\varphi,
\label{eq:diff1}\\
\mathrm{d}\phi_\text{s}&=\frac{\partial\Phi^{(n)}}{\partial\rho}\frac{\partial\mathrm{P}^{(n)}}{\partial r_\text{s}}\,\mathrm{d}r_\text{s}
+\left(\frac{\partial\Phi^{(n)}}{\partial\rho}\frac{\partial\mathrm{P}^{(n)}}{\partial\varphi}
+\frac{\partial\Phi^{(n)}}{\partial\varphi}\right)\mathrm{d}\varphi,
\label{eq:diff2}
\end{align}
\end{subequations}
where the superscripts of $R^{(n)}$ and $\Phi^{(n)}$ are understood as in
Eq.~\eqref{eq:RPhi}. Because of Eq.~\eqref{eq:identity}, we have
\begin{equation}
\frac{\partial R^{(n)}}{\partial\rho}\frac{\partial\mathrm{P}^{(n)}}{\partial r_\text{s}}=1,
\qquad
\frac{\partial R^{(n)}}{\partial\rho}\frac{\partial\mathrm{P}^{(n)}}{\partial\varphi}
+\frac{\partial R^{(n)}}{\partial\varphi}=0.
\label{eq:constraints}
\end{equation}
Substituting Eq.~\eqref{eq:constraints} into Eq.~\eqref{eq:diff2} yields
\begin{equation}
\mathrm{d}\phi_\text{s}=\left(\frac{\partial R^{(n)}}{\partial\rho}\right)^{-1}
\left[\frac{\partial\Phi^{(n)}}{\partial\rho}\,\mathrm{d}r_\text{s}
+\left|\frac{\partial\left(R^{(n)},\Phi^{(n)}\right)}{\partial(\rho,\varphi)} \right|\,\mathrm{d}\varphi\right],
\label{eq:dphi}
\end{equation}
and hence
\begin{equation}
\left|\frac{\partial\left(R^{(n)},\Phi^{(n)}\right)}{\partial(\rho,\varphi)}\right|
=\frac{\partial R^{(n)}}{\partial\rho}
\left.\frac{\partial\Phi^{(n)}\!\left(\mathrm{P}^{(n)}(r_\text{s},\theta_\text{s},r_\text{o},\theta_\text{o};\varphi),\varphi\right)}{\partial\varphi}\right|_{r_\text{s}}.
\label{eq:jac}
\end{equation}
In the standard framework of gravitational lensing, the caustics are determined by the vanishing of the Jacobian, which corresponds to $\Sigma\to\infty$. In the present setting,
this is equivalent to the vanishing of the last factor of
Eq.~\eqref{eq:jac}. Using Eqs.~\eqref{eq:Ir}, \eqref{eq:rhoeq}, and
\eqref{eq:jac}, the quantity $\Sigma$ of Eq.~\eqref{eq:invSigma} becomes
\begin{equation}
\Sigma=\frac{\rho}{r_\text{s}\sqrt{R(r_\text{s})}}
\left|\left(\frac{\partial I_r}{\partial\rho}
-\frac{\partial G_\theta^{(\text{eop},n)}}{\partial\rho}\right)
\left.\frac{\partial\Phi^{(n)}\!\left(\mathrm{P}^{(n)}(r_\text{s},\theta_\text{s},r_\text{o},\theta_\text{o};\varphi),\varphi\right)}{\partial\varphi}\right|_{r_\text{s}}\right|^{-1}.
\label{eq:Sigma}
\end{equation}
Eq.~\eqref{eq:Sigma} is the main result of this study. For a spherical black hole, the magnification of an equatorial source can be written in the same form \cite{1992A257594B,1994ApJ...425...63B,Zhu:2024vxw}, with the last factor equal to  $(1-\sin^2\theta_\text{o}\cos^2\varphi)/\cos\theta_\text{o}$ \cite{1992A257594B,Zhu:2026jom}. Our result provides its analytic generalization to a rotating spacetime. In both spherical and rotating cases, the magnification $\Sigma$ factorizes into two factors. One originates from the derivative with respect to $\rho$ and the other from the derivative with respect to $\varphi$.

\section{Caustic structure from the morphology of the azimuthal phase}
\label{sec:caustics}

Based on Eq.~\eqref{eq:Sigma}, in this section we present specific
examples of the caustic curves, of the critical curves, and of the magnification map
over the equatorial source plane. We show that the correspondence between the
multiple roots of the EOP and the caustic structure is a natural one. For brevity,
we denote $\phi_\text{s}^{(n)}(\varphi)\equiv\Phi^{(n)}_{r_\text{o},\theta_\text{o},\theta_\text{s},\phi_\text{o}}\!\left(\mathrm{P}^{(n)}(r_\text{s},\theta_\text{s},r_\text{o},\theta_\text{o};\varphi),\varphi\right)$ in the remainder of this section.

\subsection{Caustic curves and critical curves}
\label{sec:caustic}

According to Eq.~\eqref{eq:Sigma}, the caustic structure follows from setting
the Jacobian to zero, which reduces to the single condition
\begin{equation}
\frac{\partial\phi_\text{s}^{(n)}}{\partial\varphi}\equiv
\frac{\partial\Phi^{(n)}_{r_\text{o},\theta_\text{o},\theta_\text{s},\phi_\text{o}}\!\left(\mathrm{P}^{(n)}(r_\text{s},\theta_\text{s},r_\text{o},\theta_\text{o};\varphi),\varphi\right)}{\partial\varphi}=0.
\label{eq:causticcond}
\end{equation}
Because $\phi_\text{s}^{(n)}(\varphi)$ is a one-dimensional function, the caustic condition
reduces to an extremum condition. Before presenting the
specific cases, we clarify our terminology. In this study, the term ``critical curve'' refers only to the location on the image plane at which the Jacobian of the lens equation, Eq.~\eqref{eq:invSigma}, vanishes. It is not another name for the shadow boundary as used in Refs.~\cite{Gralla:2019drh,Cardenas-Avendano:2022csp}.

\begin{figure}[htbp]
\includegraphics[width=\linewidth]{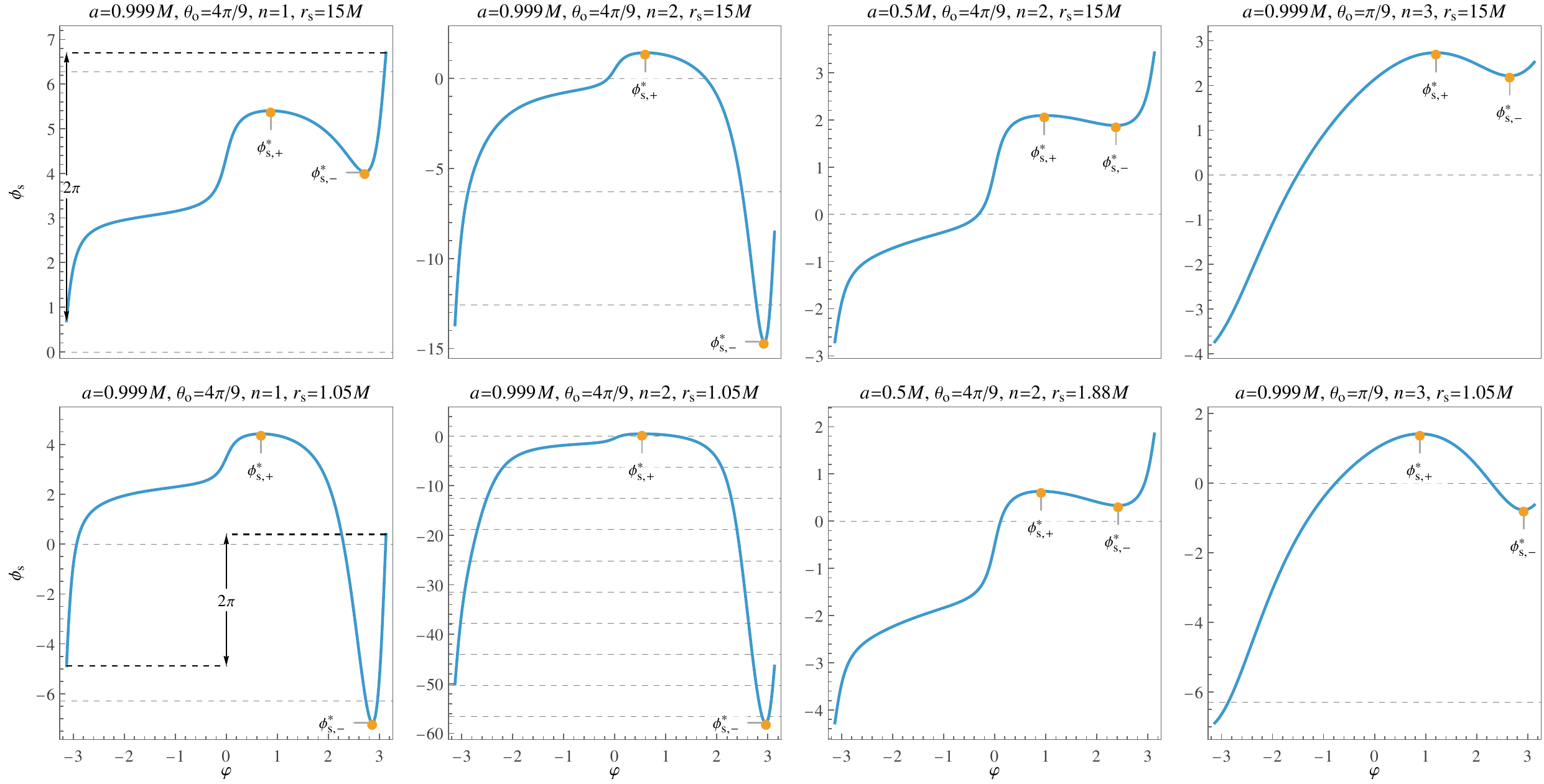}
\caption{\label{fig:landscape}The azimuthal phase $\phi_\text{s}^{(n)}(\varphi)$ as a function of the screen azimuth $\varphi$, for several
values of the spin $a$, of the inclination $\theta_\text{o}$, of the image order $n$, and of the source radius $r_\text{s}$. The dots mark the
local extrema $\phi^*_{\text{s},+}$ and $\phi^*_{\text{s},-}$, which are the caustics.
In contrast to Fig.~\ref{fig:phi}, the discontinuity at $\varphi=\pm\pi/2$ has been removed by the continuation discussed in Appendix~\ref{app:continuation}.}
\end{figure}

Fig.~\ref{fig:landscape} shows the azimuthal phase $\phi_\text{s}^{(n)}(\varphi)$ for selected values of the black hole spin and of the emitter--observer configuration. The points $\phi_{\text{s},\pm}^\ast$ are the local extrema of the azimuthal phase $\phi_\text{s}^{(n)}(\varphi)$, and they are the caustics. The critical points, denoted by $\varphi_\pm^\ast$, are the solutions of Eq.~\eqref{eq:causticcond} for $\varphi$. 
Identifying the caustics with the local extrema, and the critical points with the positions of these extrema, establishes the correspondence between the EOP and the caustic structure.
Between two consecutive extrema, the map $\varphi\mapsto\phi_\text{s}$ is one-to-one, so a source at a given $\phi_\text{s}$ has one image there. A pair of images is created or destroyed as $\phi_\text{s}$ crosses an extremum $\phi_{\text{s},\pm}^{\ast}$. The positions of the extrema vary with the spin, the inclination angle, the image order, and the source radius, which is why the caustics are not universal. In addition, the azimuthal phase satisfies the universal relation $\phi_\text{s}^{(n)}(-\pi)-\phi_\text{s}^{(n)}(\pi)=2\pi$. 

\begin{figure}[htbp]
\includegraphics[width=\linewidth]{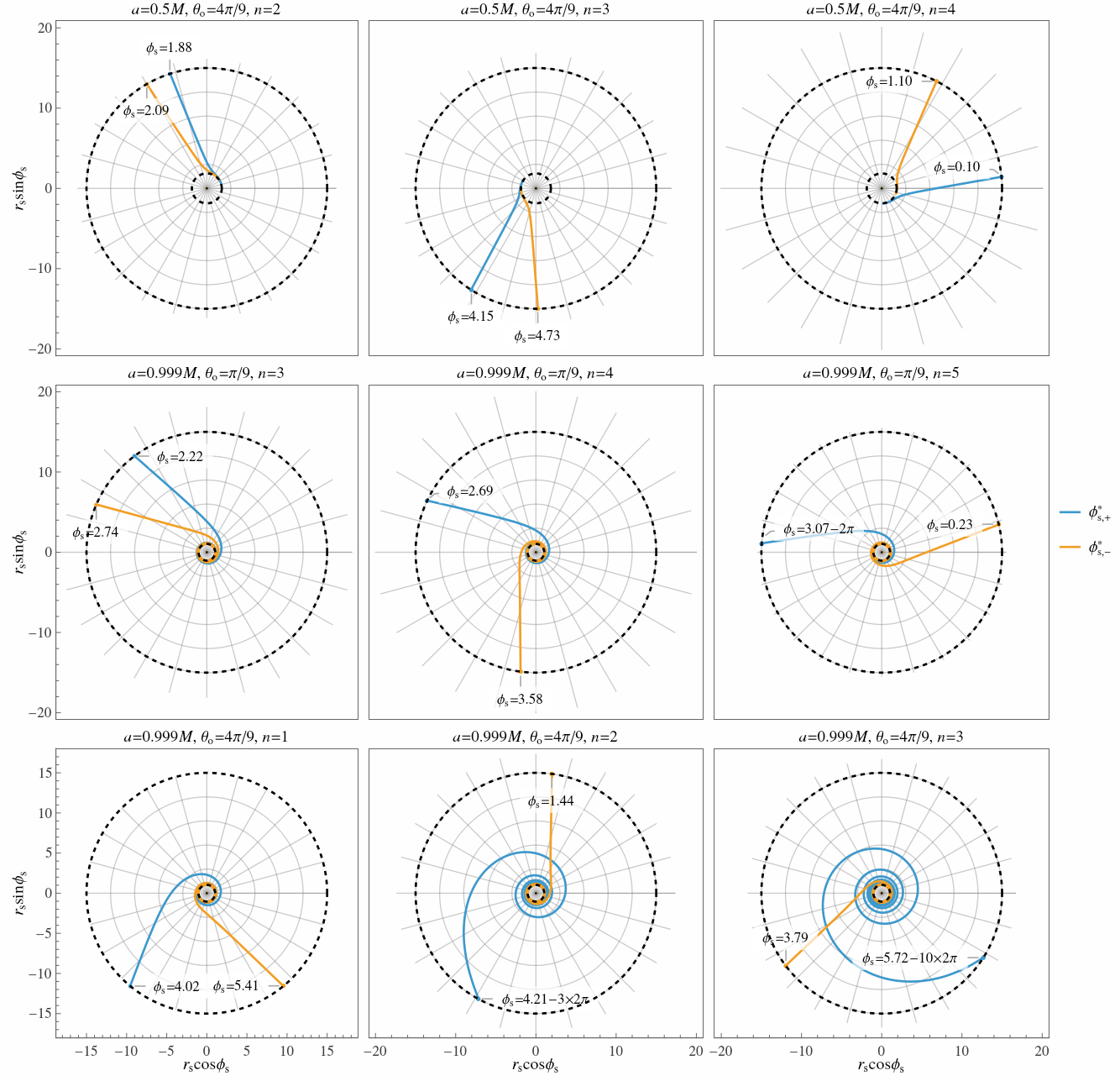}
\caption{\label{fig:caustic}Caustic curves on the equatorial source plane,
plotted in the coordinates $(r_\text{s}\cos\phi_\text{s},r_\text{s}\sin\phi_\text{s})$,
for selected spin, inclination angle, and image order. The dashed circles
mark the reference radii $r_\text{s}=15M$ and the horizon radius. The caustics of a near-horizon source are seen to wrap repeatedly around the black hole.}
\end{figure}

Fig.~\ref{fig:caustic} shows the caustic curves on the
equatorial source plane for given spin, inclination angle, and image order. Each curve is plotted according to the local extrema
$\phi^*_{\text{s},\pm}$ as the source radius varies. For a slowly rotating black hole and a low image order, the caustics reduce to straight radial lines at large source radius. As the spin and the image order increase, they wind repeatedly around the centre when approaching the horizon. Both the extent of a caustic and its winding therefore grow with the spin and with the image order.
\begin{figure}[htbp]
\includegraphics[width=\linewidth]{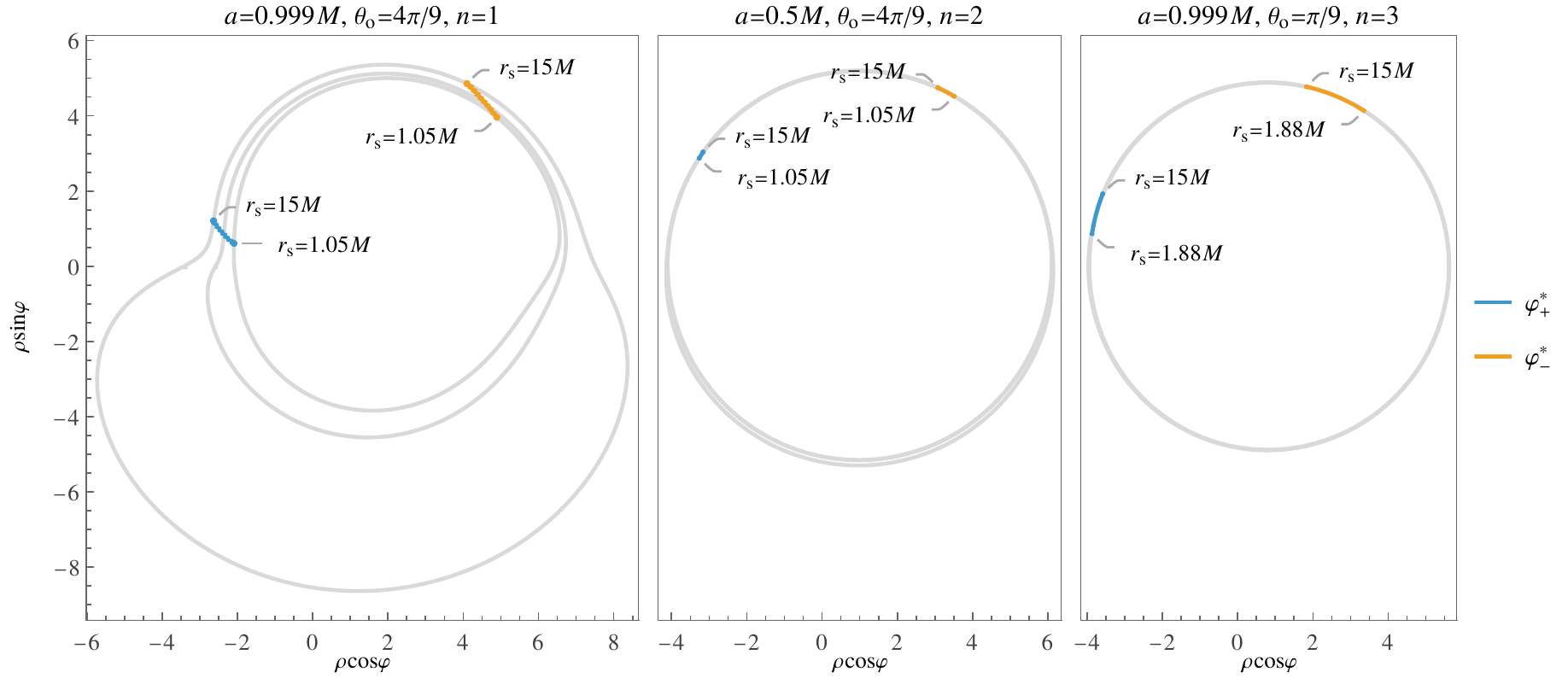}
\caption{\label{fig:critical}  Critical curves (coloured segments) that
correspond to the caustics of Fig.~\ref{fig:caustic}. The grey curves are the images of circles around the black hole that correspond
to the source radii $r_\text{s}=15M$, $r_\text{s}=2M$, and $r_\text{s}=1.05M$.}
\end{figure}
Fig.~\ref{fig:critical} shows the critical curves of the image plane that
correspond to the caustics of Fig.~\ref{fig:caustic}. The coloured segments are the critical curves, which are plotted according to the critical points $\varphi_\pm^{(n)}$ as the source radius varies. These critical curves separate the image of the equatorial source into three parts, and they are elongated as the spin or the image order increases.

\subsection{Magnification}
\label{sec:magnification}

\begin{figure}[htbp]
\includegraphics[width=\linewidth]{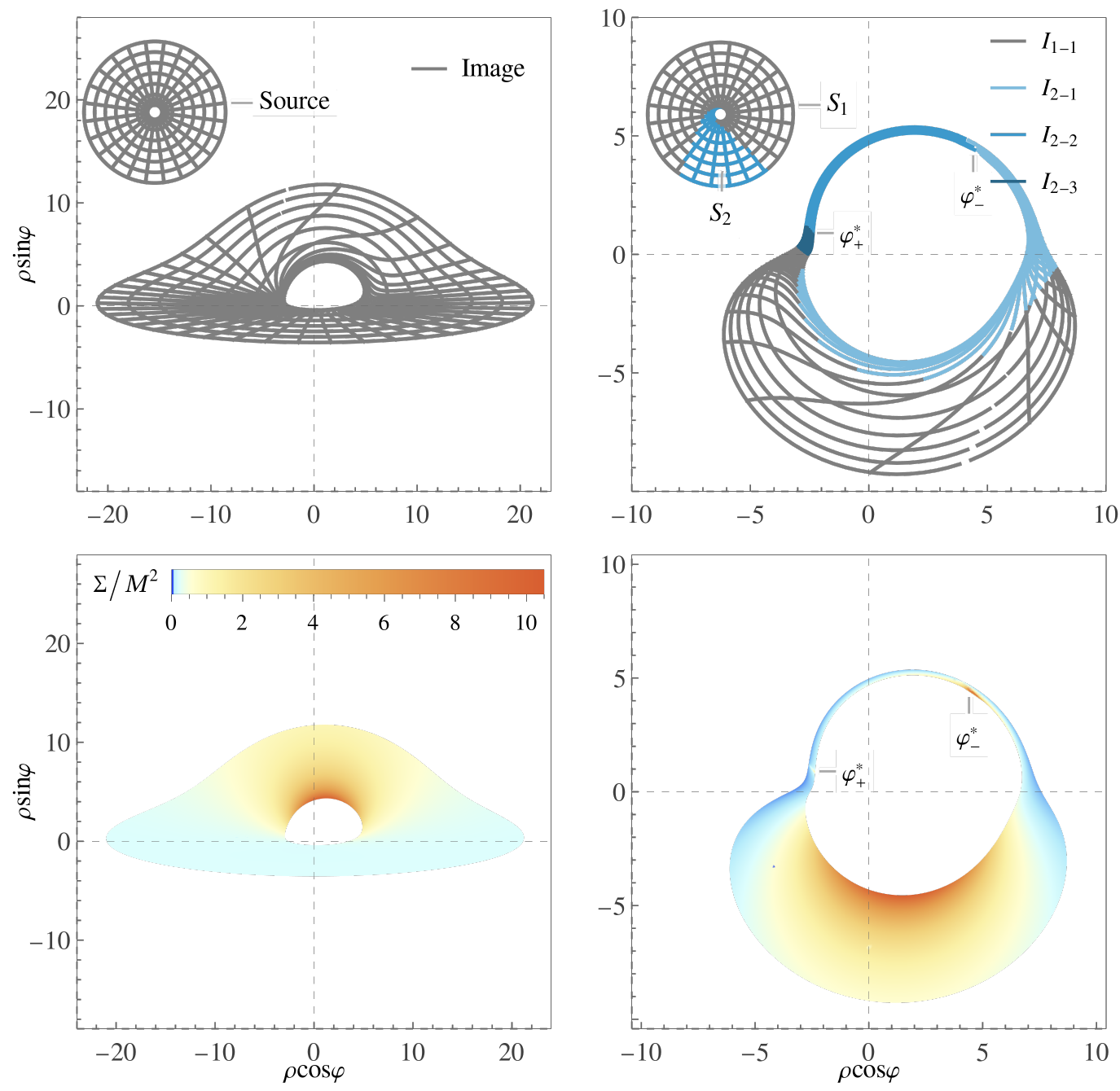}
\caption{\label{fig:grid}Upper panels: the discrete grid drawn on the equatorial disk and
its image in the image plane for the direct image $n=0$ and the subring $n=1$. Lower panels: the magnification map of $\Sigma$ for the cases of the upper panels, calculated with Eq.~\eqref{eq:Sigma}. The curves $\varphi^\ast_\pm$ represent the critical curves, as illustrated in Fig.~\ref{fig:critical}. Here, we set $a=0.999M$ and $\theta_\text{o}=4\pi/9$.}
\end{figure}

The magnification quantifies the ratio of surface elements between the image
plane and the source plane \cite{Schneider:1992bmb}. It can be measured directly by simulating the image of a thin accretion disk that carries a discrete grid, as in the upper panels of Fig.~\ref{fig:grid}. The deformation of the grid measures the Jacobian. In the upper-right panel, the disk region $S_1$, shown in grey, is mapped one-to-one onto the image $I_{1-1}$, whereas the disk region $S_2$, shown in colours, is mapped onto the three images $I_{3-1}$, $I_{3-2}$, and $I_{3-3}$. It is the caustics that map part of a ring-like source onto several images.

Without an analytic formula, however, the magnification must be obtained by measuring the area of the image-plane cells numerically. This is computationally demanding and is inaccurate near the critical curves or the horizon, where the cells are strongly distorted. A semi-analytic formula of the kind given in Eq.~\eqref{eq:Sigma} is therefore required. In Eq.~\eqref{eq:Sigma}, the derivatives $\partial I_r/\partial\rho$ and $\partial G_\theta^{(\text{eop},n)}/\partial\rho$ are analytic for the Kerr metric. The function $\phi_\text{s}^{(n)}(\varphi)$ is one-dimensional and semi-analytic, since only $\mathrm{P}^{(n)}$ of Eq.~\eqref{eq:P} is determined numerically. Using Eq.~\eqref{eq:Sigma}, we have computed the magnification of an equatorial source over the image plane. The result is shown in the lower panels of Fig.~\ref{fig:grid}. In these maps, the critical curve appears as the location at which the magnification is sharply enhanced, marked $\varphi^\ast_\pm$.

\begin{figure}[htbp]
\includegraphics[width=\linewidth]{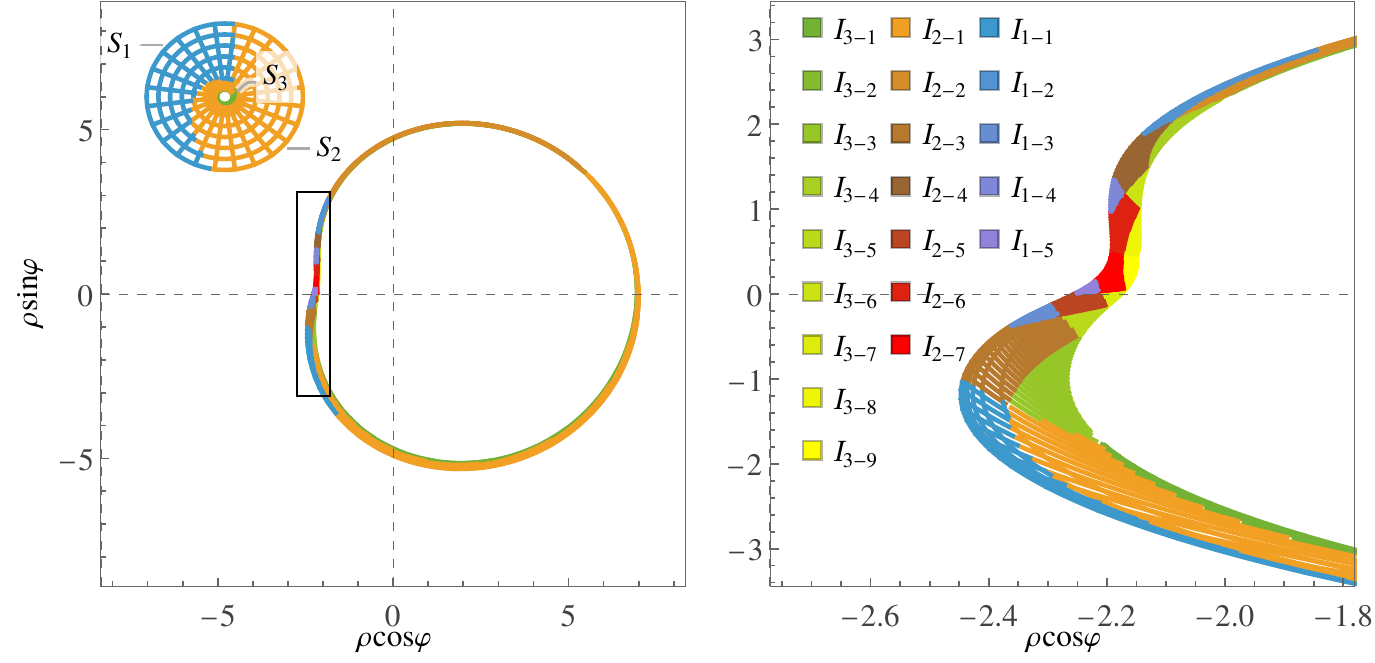}
\caption{\label{fig:grid2}Left panel: the discrete grid drawn on the equatorial disk and its
image in the image plane for the subring $n=2$. Right panel: an enlargement of the region marked by the rectangle, where
the multiple images produced by the caustics are visible. Here, we set $a=0.999M$ and $\theta_\text{o}=4\pi/9$.}
\end{figure}

Fig.~\ref{fig:grid2} shows the equatorial disk in a discrete grid and its image at the image order $n=2$. The disk is divided into the three regions $S_1$, $S_2$, and $S_3$, drawn in different colours. The numbers of their images are $5$, $7$, and $9$, respectively, as shown in the right panel of Fig.~\ref{fig:grid2}. Towards the inner side of the images, the multiplicity increases from $5$ to $9$. Since these multiple images originate from a single region, their correlated structure on the subring is expected to carry information about the spacetime geometry.

\subsection{Multiplicity of the images}
\label{sec:multiplicity}

For a given image order $n$, we can solve Eq.~\eqref{eq:phieq} exactly for the
multiple values of $\varphi$ at a specified $\phi_\text{s}$, as illustrated in Fig.~\ref{fig:landscape}. Once the solutions are found, the multiplicity of the images follows by counting. To a good approximation, the
multiplicity can also be estimated from the span of the azimuthal phase,
\begin{equation}
N=2\left\lfloor\frac{\Delta\phi_n}{2\pi}\right\rfloor+1,
\qquad
\Delta\phi_n=\phi^{(n)}_{\text{s},\max}(\varphi)-\phi^{(n)}_{\text{s},\min}(\varphi).
\label{eq:N}
\end{equation}
Note that $\Delta\phi_n$ need not equal $\phi^\ast_{\text{s},+}-\phi^\ast_{\text{s},-}$.
Eq.~\eqref{eq:N} holds because every complete winding of $\phi_\text{s}^{(n)}$ adds a pair of images,
with a single image remaining when the span is smaller than $2\pi$. 

In
Table~\ref{tab:multiplicity}, we compare the two determinations of $N$ for
selected values of the spin and of the source azimuth.
\begin{table}[htbp]
\caption{\label{tab:multiplicity}Multiplicity $N$ of the images of a point
source on the equatorial plane, for selected image orders and spins. The first six
columns give the exact count obtained by solving Eq.~\eqref{eq:phieq} at the
source azimuths $\phi_\text{s}=5$ and $\phi_\text{s}=2$, and the last three
give the estimate of Eq.~\eqref{eq:N}. Within each group, the observer is quasi-equatorial, $\theta_\text{o}=\pi/2-\pi/40$.}
\begin{ruledtabular}
\begin{tabular}{c@{\hspace{3em}} ccc @{\hspace{0.8em}} ccc @{\hspace{2.2em}} ccc}
&\multicolumn{6}{c}{Exact count} & \multicolumn{3}{c}{Estimate} \\
& \multicolumn{3}{c}{$\phi_\text{s}=5$} & \multicolumn{3}{c}{$\phi_\text{s}=2$} & \multicolumn{3}{c}{Eq.~\eqref{eq:N}} \\
$n\backslash a$ & $0.99$ & $0.999$ & $0.9999$ &
$0.99$ & $0.999$ & $0.9999$ &
$0.99$ & $0.999$ & $0.9999$ \\
\hline
0 & 1 & 1 & 1 & 1 & 1 & 1 & 1 & 1 & 1 \\
1 & 1 & 1 & 1 & 3 & 3 & 3 & 1 & 1 & 1 \\
2 & 5 & 7 & 7 & 5 & 7 & 7 & 5 & 7 & 7 \\
3 & 11 & 23 & 35 & 11 & 23 & 35 & 11 & 23 & 35 \\
4 & 17 & 45 & 95 & 17 & 43 & 95 & 17 & 43 & 93 \\
5 & 23 & 65 & 161 & 23 & 65 & 163 & 23 & 63 & 161 \\
6 & 29 & 85 & 229 & 29 & 87 & 231 & 29 & 85 & 229 \\
\end{tabular}
\end{ruledtabular}
\end{table}
The exact count
is odd at every order, as it must be, because images are created and destroyed
in pairs. At a given image order, it grows with the spin, and the growth is steep, from
11 images at $n=3$ and $a=0.99M$ to 35 at $a=0.9999M$. The
estimate reproduces the count to within two images. The origin of the difference is the discreteness of Eq.~\eqref{eq:N}, which counts complete windings only.

\begin{figure}[htbp]
\includegraphics[width=\linewidth]{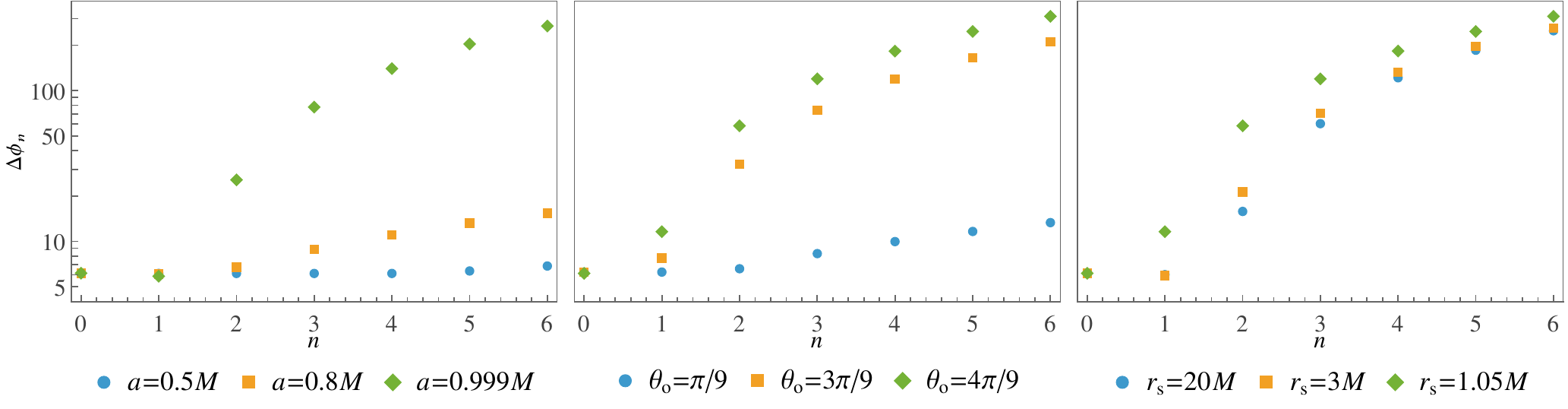}
\caption{\label{fig:delta}Span $\Delta\phi_n$ of the azimuthal phase $\phi_\text{s}^{(n)}(\varphi)$ as a function of the image order $n$, for different spins (left
panel), inclinations (middle panel), and source radii (right panel). The
vertical axis is logarithmic.}
\end{figure}

Fig.~\ref{fig:delta} shows the span $\Delta\phi_n$ of the azimuthal phase $\phi_\text{s}^{(n)}$ as a function of the image order, for selected spins, inclinations, and source
radii. At high orders, the span grows linearly
with $n$ in all three panels, and the slope depends on these parameters. Through
Eq.~\eqref{eq:N}, this produces the linear growth of the number of
images. Of the three parameters, the spin
produces the strongest dependence. For $a=0.5M$, the span stays below ten at all orders
shown, whereas for $a=0.999M$ it exceeds two hundred at $n=6$. The dependence
on the source radius is the weakest of the three at large radii, but it becomes strong once the source lies within a few gravitational radii of the horizon.

\section{Conclusions and discussion}
\label{sec:conclusions}

We have developed a new implementation of the Kerr emitter--observer problem
and have used it to formulate the caustic structure of the equatorial source. In
this implementation, the two-dimensional root finding that is customary in
the EOP is replaced by two successive one-dimensional problems. On this basis, we have established a correspondence between the caustic structure of strong gravitational lensing and the
image multiplicity of the EOP. With this implementation, we have obtained semi-analytic results for the caustic curves, for the critical curves, and for
the magnification of an equatorial source around a rotating black hole.

The semi-analytic magnification in Eq.~\eqref{eq:Sigma} is useful precisely where grid-based methods fail,
that is, where the image is strongly distorted near the horizon, as
Fig.~\ref{fig:grid} illustrates. The caustic structure of a near-horizon source is
qualitatively different from that of a distant source. The caustics wrap
several times around the black hole and their winding grows with the spin and
with the image order, as shown in Figs.~\ref{fig:caustic} and~\ref{fig:critical}.
In addition, our EOP formalism finds all the images. It
therefore provides a complete computational approach that differs from the
shooting method used in EOP codes \cite{Yang:2013wya}. By exploring the parameter space, we
have observed a linear dependence of the multiplicity $N$ on the image order
$n$. The slope can be modulated by the spin, the inclination, and the
distance of the source from the black hole (Fig.~\ref{fig:delta} and
Table~\ref{tab:multiplicity}). This is consistent with the standard description of gravitational lensing \cite{Bozza:2008mi}. Because the estimate in Eq.~\eqref{eq:N} reproduces the exact count to within two images, the multiplicity
can be predicted from a single winding number of the azimuthal phase.

In principle, the multiplicity in the EOP emerges from the root-finding procedure
and is independent of how the source plane and the Jacobian are defined. The
coincidence of the caustics with the local extrema of the azimuthal phase $\phi_\text{s}^{(n)}(\varphi)$ is
therefore not guaranteed by construction. Our conclusions have to be
restricted to the setup used here, which is an equatorial
source. In particular, it is not clear how the standard formulation of
gravitational lensing with a source plane $(\theta_\text{s},\phi_\text{s})$ at
large distance is related to the EOP for a near-horizon source. This is a
question for future work.
For an equatorial source, we have found that the images of a fixed order are
located on a common closed curve in the image plane, as shown in
Fig.~\ref{fig:image}. This is not a general property of the EOP. It holds
only because we restrict ourselves to the regime
$\sin\theta_\text{s}>\sin\theta_\text{o}$, which is the regime of a source on
the equatorial plane seen by an inclined observer. In this
regime, Eq.~\eqref{eq:P} has a unique solution.

\begin{acknowledgments}
This work is supported by the National Natural Science Foundation of China
under Grants No.~12305073 and No.~12347101. The author thanks Prof.~Xin Li
for useful discussions.
\end{acknowledgments}

\appendix

\section{Reverse ray tracing}
\label{app:reversetrace}

In the implementation of the EOP presented in Sec.~\ref{sec:eopnew}, we
introduced $G_\ast^{(n)}$ to define the image order correctly, with
$G^{(n)}$ serving as an intermediate quantity. $G^{(n)}$ is the quantity most
frequently used in reverse ray tracing, in which a ray is traced
backwards from the observer to the source. The equations for the reverse
tracing are
\begin{subequations}
    \begin{align}
\tau&=G_{\theta}^{(n)}(\theta_\text{o},\theta;\rho,\varphi,\theta_\text{o}),
\label{eq:rev1}\\
r&=I_r^{-1}\!\left(\tau;\rho,\varphi,\theta_\text{o},r_\text{o}\right),
\label{eq:rev2}\\
\phi&=\phi_\text{o}-aI_{\phi}(r_\text{o},r;\rho,\varphi,\theta_\text{o})
+\rho\sin\theta_\text{o}\cos\varphi\,G_{\phi}^{(n)}(\theta_\text{o},\theta;\rho,\varphi),
\label{eq:rev3}
\end{align}
\end{subequations}
where $\tau$ is the Mino time and $I_r^{-1}$ is the inverse of the radial
integral.

\begin{figure}[htbp]
\includegraphics[width=\linewidth]{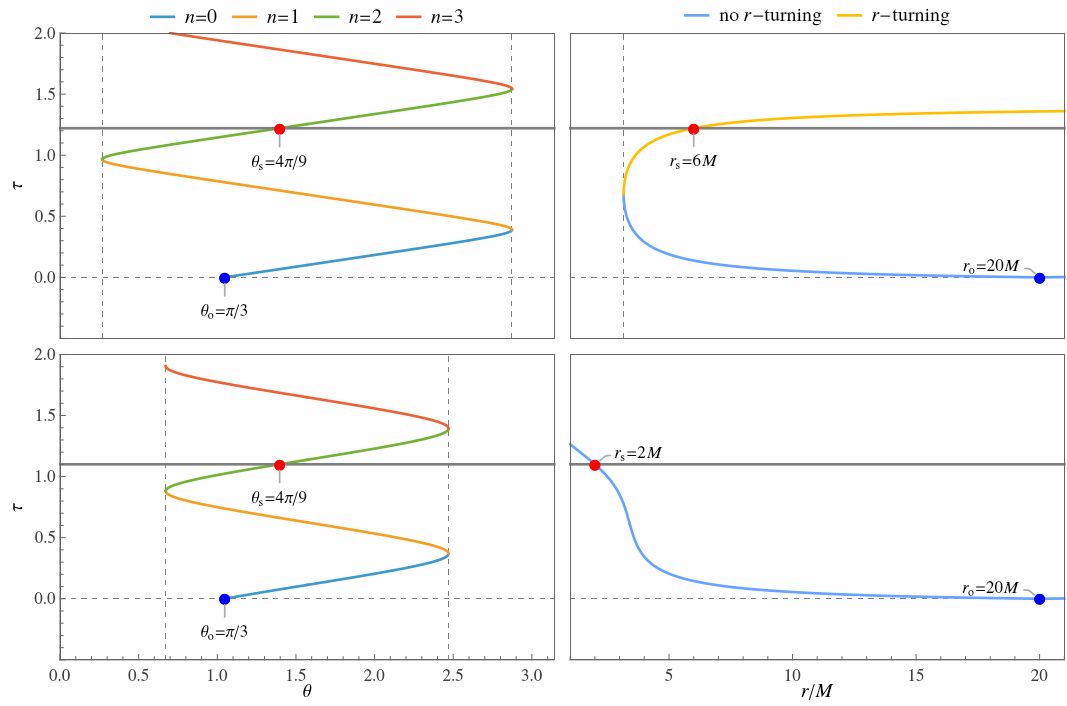}
\caption{\label{fig:path} Schematic diagrams for the reverse ray tracing.
Left panels: the Mino time
$\tau=G_\theta^{(n)}(\theta_\text{o},\theta;\rho,\varphi)$ as a function of the
polar angle $\theta$, for the image orders $n=0$ to $3$ and two source positions.
Right panels: the radius reconstructed from Eq.~\eqref{eq:rev2} as a function
of $\tau$, for the branches without and with a radial turning point. The red and blue dots
mark the source and the observer, respectively.}
\end{figure}

To trace the path of a light ray with a given impact parameter
$(\rho,\varphi)$ from $\mathbf{x}_\text{o}$ to $\mathbf{x}_\text{s}$, the
polar angle must run through the branches $n=0$ up to the specified $n$. The
radius must start from the ``in'' branch and end in the final state. This scheme is used for plotting the light paths in the right panels of Fig.~\ref{fig:image}. Here, we further illustrate the
procedure in Fig.~\ref{fig:path}.
The left panels show the
Mino time $\tau$ of Eq.~\eqref{eq:rev1} as a function of the polar angle for the
orders $n=0$ to $3$. Each branch is smooth, and consecutive branches join at
the turning points $\theta_\pm$, so the angular motion can be followed across
the oscillations without a root search. The right panels show the radius
reconstructed from Eq.~\eqref{eq:rev2}, for the branches
with and without a radial turning point. The two branches meet at the turning
point. The branch without a turning point decreases monotonically from the
observer to the source, whereas the branch with a turning point reaches a
minimum radius and then returns outward.

\section{Continuation of the azimuthal phase}
\label{app:continuation}

A possible objection to the validity of Eq.~\eqref{eq:jac} is that the
derivative of the composite function $\phi_\text{s}^{(n)}(\varphi)$ introduced in
Sec.~\ref{sec:caustics} is not defined at $\varphi=\pm\pi/2$, because of the
properties of the elliptic integral entering $G_\theta$. It can nevertheless be verified that
\begin{equation}
\Phi^{(n)}\big|_{\varphi=(\pi/2)^+}-\Phi^{(n)}\big|_{\varphi=(\pi/2)^-}=2\pi k,
\qquad k\in\mathbb{Z},
\label{eq:jump}
\end{equation}
with a similar relation at $\varphi=-\pi/2$. Because of this, and owing to the
periodicity of $\phi_\text{s}$, the two sides of $\pm\pi/2$ can be identified.
This is achieved by introducing the piecewise constant
\begin{equation}
\begin{aligned}
F_n(\varphi)=2\pi\Big[&n+\mathcal{H}\!\left(\varphi+\frac{\pi}{2}\right)\mathcal{H}\!\left(\frac{\pi}{2}-\varphi\right)
-\mathcal{H}(\theta_\text{s}-\theta_\text{o})\mathcal{H}\!\left(\varphi-\frac{\pi}{2}\right)-\mathcal{H}(\theta_\text{o}-\theta_\text{s})\mathcal{H}\!\left(-\frac{\pi}{2}-\varphi\right)\Big],
\end{aligned}
\label{eq:Fn}
\end{equation}
in terms of which the continuous version of the azimuthal phase is
\begin{equation}
\tilde{\Phi}^{(n)}\equiv\Phi^{(n)}+F_n(\varphi).
\label{eq:Phitilde}
\end{equation}
The integer $k$ of Eq.~\eqref{eq:jump} is then absorbed into the constant
offset that $F_n$ supplies on each of the three intervals
$(-\pi,-\pi/2)$, $(-\pi/2,\pi/2)$, and $(\pi/2,\pi)$, so that
$\tilde{\Phi}^{(n)}$ is continuous and can be differentiated across
$\varphi=\pm\pi/2$. Since $F_n$ is constant on each interval, it does not
affect the caustic condition \eqref{eq:causticcond}.

\section{Analytic form of the integrals}
\label{app:integrals}

In this appendix, we summarize the analytic integrals that are used in this study. The elliptic functions and the Carlson functions can all be found in the classical Refs.~\cite{byrd_friedman_1971,NIST:DLMF,gradshteyn_ryzhik_2014,Gralla:2019drh}.

\subsection{Roots of the angular and radial equations}

\subsubsection{The equation $R(r)=0$}

Expanding Eq.~\eqref{eq:potentials}, the radial potential is the quartic
\begin{equation}
R(r)=r^4+Ar^2+Br+C,
\label{eq:quartic}
\end{equation}
with
\begin{equation}
A=2a^2-\kappa-2a\lambda,\qquad B=2M\kappa,\qquad C=a^2\left(a^2-\kappa-2a\lambda+\lambda^2\right).
\label{eq:quartcoef}
\end{equation}
Its roots are obtained in closed form by the Ferrari--Cardano method \cite{birkhoff_maclane_1965}. Defining
\begin{align}
\Delta_q&=-4\left(A^2+12C\right)^3+\left(2A^3+27B^2-72AC\right)^2,
\label{eq:Deltaq}\\
\Phi_q&=2A^3+27B^2+\sqrt{\Delta_q}-72AC,
\label{eq:Phiq}\\
F_q&=\sqrt{-\frac{2A}{3}+\frac{2^{1/3}\left(A^2+12C\right)}{3\Phi_q^{1/3}}+\frac{\Phi_q^{1/3}}{3\cdot2^{1/3}}},
\label{eq:Fq}
\end{align}
the four roots are
\begin{align}
r_{1,2}&=-\frac{F_q}{2}\mp\frac{1}{2}\sqrt{-2A+\frac{2B}{F_q}-F_q^2},
\label{eq:roots12}\\
r_{3,4}&=\frac{F_q}{2}\mp\frac{1}{2}\sqrt{-2A-\frac{2B}{F_q}-F_q^2},
\label{eq:roots34}
\end{align}
ordered so that real roots come first in ascending order, followed by the
complex pair when only two roots are real. Depending on the sign of
$\Delta_q$ and on the signs of the radicands, three cases arise, namely four
real roots, two real roots with one complex conjugate pair, and two complex
conjugate pairs. For the impact parameters of interest here, the first two cases occur. The third admits no turning point and plays no role in the branch
structure discussed in Sec.~\ref{sec:branch}. When all four roots are real,
$r_4$ is the largest one and plays the role of the radial turning point in
Eq.~\eqref{eq:turning}.

\subsubsection{The equation $\Theta(\theta)=0$}

The angular potential vanishes when
\begin{equation}
a^2\sin^4\theta-(\kappa+2a\lambda)\sin^2\theta+\lambda^2=0,
\label{eq:thetaquad}
\end{equation}
so that the two roots entering the factorization
\eqref{eq:factored} are
\begin{equation}
u_\pm=\frac{\kappa+2a\lambda\pm\sqrt{\kappa\left(\kappa+4a\lambda\right)}}{2a^2}.
\label{eq:upm}
\end{equation}
Rays that reach the equatorial plane require $u_+>1$ and $0<u_-<1$, so that
the polar motion oscillates between
$\theta_-=\arcsin\sqrt{u_-}$ and $\theta_+=\pi-\arcsin\sqrt{u_-}$. When the
radicand $\kappa(\kappa+4a\lambda)$ is negative, or when no root falls in
$(0,1)$, the light ray never reaches the equatorial plane and belongs to the class
of vortical geodesics \cite{Gralla:2019drh}.

\subsection{Transfer functions}

\subsubsection{Radial integral}

The radial integral in Eq.~\eqref{eq:Ir} is an elliptic integral of the first kind
with four roots. It is most conveniently evaluated in the Carlson symmetric
form \cite{NIST:DLMF}. For the branch without a turning point,
\begin{equation}
I_r^\text{in}=2R_F\!\left(U_{12}^2,U_{13}^2,U_{23}^2\right),
\label{eq:IrCarlson}
\end{equation}
where
\begin{align}
U_{12}&=\frac{\sqrt{r_\text{o}-r_1}\sqrt{r_\text{o}-r_2}\sqrt{r_\text{s}-r_3}\sqrt{r_\text{s}-r_4}
+\sqrt{r_\text{s}-r_1}\sqrt{r_\text{s}-r_2}\sqrt{r_\text{o}-r_3}\sqrt{r_\text{o}-r_4}}{r_\text{o}-r_\text{s}},
\label{eq:U12}\\
U_{13}&=\frac{\sqrt{r_\text{o}-r_1}\sqrt{r_\text{o}-r_3}\sqrt{r_\text{s}-r_2}\sqrt{r_\text{s}-r_4}
+\sqrt{r_\text{s}-r_1}\sqrt{r_\text{s}-r_3}\sqrt{r_\text{o}-r_2}\sqrt{r_\text{o}-r_4}}{r_\text{o}-r_\text{s}},
\label{eq:U13}\\
U_{23}&=\frac{\sqrt{r_\text{o}-r_2}\sqrt{r_\text{o}-r_3}\sqrt{r_\text{s}-r_1}\sqrt{r_\text{s}-r_4}
+\sqrt{r_\text{s}-r_2}\sqrt{r_\text{s}-r_3}\sqrt{r_\text{o}-r_1}\sqrt{r_\text{o}-r_4}}{r_\text{o}-r_\text{s}}.
\label{eq:U23}
\end{align}
For the branch with a turning point at $r_4$, the integral is the sum of two
pieces,
\begin{equation}
I_r^\text{out}=2R_F\!\left(U_{12}^{(\text{t})2},U_{13}^{(\text{t})2},U_{23}^{(\text{t})2}\right)\Big|_{r=r_\text{s}}
+2R_F\!\left(U_{12}^{(\text{t})2},U_{13}^{(\text{t})2},U_{23}^{(\text{t})2}\right)\Big|_{r=r_\text{o}},
\label{eq:IrOut}
\end{equation}
where the endpoint $r_4$ replaces $r_\text{o}$ in the expressions
\eqref{eq:U12}--\eqref{eq:U23}, that is,
\begin{equation}
U_{12}^{(\text{t})}=\frac{\sqrt{r_4-r_1}\sqrt{r_4-r_2}\sqrt{r-r_3}\sqrt{r-r_4}}{r_4-r},
\label{eq:U12t}
\end{equation}
and $U_{13}^{(\text{t})}$ and $U_{23}^{(\text{t})}$ follow by the corresponding
cyclic permutations of the indices. The inverse of the radial integral, which
is needed in Eq.~\eqref{eq:Rn}, is obtained in closed form in terms of Jacobi
elliptic functions. For four real roots and $r\ge r_4$,
\begin{equation}
I_r^{(4)}(r)=\frac{2}{\sqrt{(r_4-r_2)(r_3-r_1)}}\,
F\!\left(\arcsin\sqrt{\frac{(r_3-r_1)(r-r_4)}{(r_4-r_1)(r-r_3)}}\,\Big|\,m\right),
\label{eq:IrF}
\end{equation}
with
\begin{equation}
m=\frac{(r_3-r_2)(r_4-r_1)}{(r_4-r_2)(r_3-r_1)},
\label{eq:m}
\end{equation}
whose inverse is
\begin{equation}
R^{(4)}(u)=\frac{s\,(r_4-r_1)\,r_3-r_4\,(r_3-r_1)}{s\,(r_4-r_1)-(r_3-r_1)},
\qquad s=\mathrm{sn}^2\!\left(\frac{u\,(r_4-r_2)(r_3-r_1)}{2}\,\Big|\,m\right).
\label{eq:Rinv}
\end{equation}
$F(\phi|m)$ and $\mathrm{sn}(u|m)$ are the incomplete elliptic integral
of the first kind and the Jacobi elliptic sine, respectively. For the branch with a turning
point, the source radius is recovered from
$I_r^{(4)}(r_\text{s})=I_r^\text{out}-I_r^{(4)}(r_\text{o})$.

The azimuthal integral in Eq.~\eqref{eq:Iphi} contains the two horizon poles
$r_\pm=M\pm\sqrt{M^2-a^2}$. Writing $\Delta=(r-r_+)(r-r_-)$ and decomposing
the integrand, one obtains
\begin{equation}
I_{\phi}=\frac{a}{r_+-r_-}\left[\left(2Mr_+-a\lambda\right)J(r_+)-\left(2Mr_--a\lambda\right)J(r_-)\right],
\label{eq:IphiJ}
\end{equation}
with the integrals
\begin{equation}
J(p)=\int_{r_\text{s}}^{r_\text{o}}\frac{\mathrm{d}r}{(r-p)\sqrt{R(r)}}.
\label{eq:J}
\end{equation}
Since the poles lie inside the horizon and therefore outside the integration
range, $J$ is regular. Its numerical value is nevertheless the difference of
two large quantities, and its evaluation requires the subtraction
\begin{multline}
J(p)=\frac{1}{p-r_1}\Bigg[\frac{2}{3}\,(r_2-r_1)(r_3-r_1)\frac{r_4-r_1}{p-r_1}\,
R_J\!\left(U_{12}^2,U_{13}^2,U_{23}^2,U_{15}^2\right)\\
+2R_C\!\left(S_{15}^2,Q_{15}^2\right)-2R_F\!\left(U_{12}^2,U_{13}^2,U_{23}^2\right)\Bigg],
\label{eq:Jcarlson}
\end{multline}
where $R_J$ and $R_C$ are the Carlson integrals of the third kind and of the
degenerate kind, and
\begin{align}
U_{15}^2&=U_{12}^2-\frac{(r_3-r_1)(r_4-r_1)(p-r_2)}{p-r_1},
\label{eq:U15}\\
S_{15}&=\frac{1}{r_\text{o}-r_\text{s}}\left[\sqrt{r_\text{o}-r_2}\sqrt{r_\text{o}-r_3}\frac{\sqrt{r_\text{o}-r_4}}{\sqrt{r_\text{o}-r_1}}\,(r_\text{s}-p)
+\sqrt{r_\text{s}-r_2}\sqrt{r_\text{s}-r_3}\frac{\sqrt{r_\text{s}-r_4}}{\sqrt{r_\text{s}-r_1}}\,(r_\text{o}-p)\right],
\label{eq:S15}\\
Q_{15}^2&=\left(\frac{\sqrt{r_\text{o}-p}\sqrt{r_\text{s}-p}}{\sqrt{r_\text{o}-r_1}\sqrt{r_\text{s}-r_1}}\right)^2U_{15}^2.
\label{eq:Q15}
\end{align}

\subsubsection{Angular integral}

The angular integrals in Eq.~\eqref{eq:transfers} reduce to Legendre normal forms. With
$u_\pm$ given by Eq.~\eqref{eq:upm} and
\begin{equation}
z(\theta)\equiv\frac{\cos\theta}{\sqrt{1-u_-}},
\qquad
m_\theta\equiv\frac{1-u_-}{u_+-u_-},
\label{eq:ztheta}
\end{equation}
one has
\begin{align}
G_\theta(\theta)&=\frac{1}{a\sqrt{u_+-u_-}}\,F\!\left(\arccos z\,\big|\,m_\theta\right),
\label{eq:GthetaF}\\
G_\phi(\theta)&=\frac{1}{a\,u_-\sqrt{u_+-u_-}}\,\Pi\!\left(\frac{u_--1}{u_-};\arccos z\,\Big|\,m_\theta\right),
\label{eq:GphiPi}
\end{align}
where $F$ and $\Pi$ are the incomplete elliptic integrals of the first and
third kind. The corresponding complete integrals, which are the
contributions of a full half-orbit $\int_{\theta_-}^{\theta_+}$, are
\begin{align}
G_\theta^{+}&=\frac{2}{a\sqrt{u_+-u_-}}\,K(m_\theta),
\label{eq:GthetaK}\\
G_\phi^{+}&=\frac{2}{a\,u_-\sqrt{u_+-u_-}}\,\Pi\!\left(\frac{u_--1}{u_-}\,\Big|\,m_\theta\right),
\label{eq:GphiPicomplete}
\end{align}
with $K$ the complete elliptic integral of the first kind. Substituting
Eqs.~\eqref{eq:GthetaF}--\eqref{eq:GphiPicomplete} into Eq.~\eqref{eq:Gop} gives
the angular transfer function $G_\ast^{(\text{eop},n)}$ that enters the radial equation
\eqref{eq:rhoeq} and the lens equation \eqref{eq:lens}. Because the complete
integrals are independent of both $\theta_\text{s}$ and $\theta_\text{o}$, the
dependence on the image order is entirely contained in the integer
coefficients of Eq.~\eqref{eq:Gop}.

\bibliography{ref}

\end{document}